\documentclass[conference, 10 pt]{IEEEtran}
\usepackage[letterpaper,margin = 1in]{geometry}
\usepackage[utf8]{inputenc}
\usepackage{cite}
\usepackage{amsmath}
\usepackage{verbatim}
\usepackage{tikz}
\usepackage{algorithmic}
\usepackage{array}
\usepackage{caption}
\usepackage{subcaption}
\usepackage{graphicx}
\usepackage{multicol}
\usepackage{dblfloatfix}
\usepackage{amsmath}
\usepackage{amssymb}
\usepackage{algorithm}
\usepackage{url}
\usepackage{comment}
\usepackage{makecell}

\title{Channel Estimation for Massive MIMO systems using Tensor Cores in GPU}
\author{\IEEEauthorblockN{Bhargav Gokalgandhi\IEEEauthorrefmark{1}, Ivan Seskar\IEEEauthorrefmark{2} \\}
\IEEEauthorblockA{WINLAB, Rutgers University\\
671, US-1, North Brunswick Township, NJ 08902 \\
Email: \IEEEauthorrefmark{1}bgokal@winlab.rutgers.edu, \IEEEauthorrefmark{2}seskar@winlab.rutgers.edu}
\thanks{This project was funded by the NSF "COSMOS" Project under grant number CNS-1827923.}
}

\IEEEoverridecommandlockouts

\begin{document}

\maketitle

\begin{abstract}
    For efficient use of Massive MIMO systems, fast and accurate channel estimation is very important. But the Large-scale antenna array presence requires high pilot overhead for high accuracy of estimation. Also, when used with software-based processing systems like CPUs and GPUs, high processing latency becomes a major issue. To reduce Pilot overhead, a Pilot transmission scheme in combination with PN Sequence correlation based channel estimation scheme is implemented. Then, to deal with the issue of high processing latency, Tensor Cores in Nvidia GPUs are used for computing the channel estimation. Experiments are performed by using Nvidia V100 GPU in the ORBIT Testbed to show the performance of the Pilot transmission scheme. By varying factors like PN sequence length, Channel Impulse Response length, number of multiplexed transmitters, and scale of MIMO, the accuracy and processing latency of Tensor Core implementation of the Channel Estimation is evaluated.
\end{abstract}

\section{Introduction}\label{sec:intro}

To effectively use Massive MIMO systems, channel characterization and Channel State Information (CSI) acquisition are extremely crucial tasks. When performing real-time Channel Sounding and Channel Estimation in practical scenarios, issues related to high pilot overhead and pilot contamination need to be addressed \cite{xu2013effect}. The Large-scale antenna systems present at transmitter as well as the receiver make it difficult to balance the pilot overhead and data transmission time within the limited channel coherence time window. Furthermore, careful design and transmission of Pilot sequences to get reliable CSI is extremely important for precoding/decoding of data. Channel Sounding techniques and architectures for CSI acquisition in a variety of Massive MIMO and mmWave scenarios have been considered \cite{bas2019real,laly2020massive,yang2018parallel,wen2016mmwave,fan2019development,zhao2017system,wang2020enabling,grcon}. Also, techniques which address issues related to pilot contamination, pilot sequence overhead, antenna array design, etc. have been extensively studied for various Massive MIMO scenarios \cite{hassan2020channel,benzaghta2021massive,hassan2021survey,shaik2021comprehensive,chen2021survey}.

Recently, consideration has been given to combining Massive MIMO systems with high-performance servers consisting of multi-core CPUs and GPUs at the back-end for baseband processing. CPUs and GPUs can provide fast processing while giving high flexibility for implementation and testing of various baseband processing tasks. 
But the Large-Scale antennas arrays at the front-end of Massive MIMO systems result in high volume of data to be processed. This makes the processing for real-time Channel Estimation extremely difficult when using CPUs and GPUs for processing.

So, an investigation into the performance of channel estimation for Massive MIMO systems when implemented in software-based processing systems like GPU is required. Especially, consideration into the trade-off between processing latency and estimation accuracy is required when implementing algorithms in real-time Massive MIMO scenarios. Recently, investigation has been done related to implementation of various channel estimation algorithms. In \cite{gokalgandhi2019accelerating}, Least Squares Channel Estimation is implemented by using GPU and its performance in terms of processing latency is shown. 
\cite{riadi2019performance} measures the performance of Least Squares Channel Estimation of Massive MIMO-OFDM systems through the error performance obtained after detection of M-QAM symbols. \cite{belgiovine2021deep,jin2019channel} investigate the use of Deep Learning with GPUs to accelerate the channel estimation process for mmWave Massive MIMO systems. An issue with above channel estimation algorithms is that they have either bad error performance at low SNRs in case of LS, or in case of Deep Learning require training and high volume of initial data to get the required high performance. So, to solve the issue of low training overhead, and to maintain high estimation accuracy and low computational complexity, we look into sequence correlation based channel estimation. The advantages and limitations of correlation based channel estimation have been investigated \cite{liu2012improved,yu2019channel}, but its implementation in a software-based processing system is yet to be completely explored. Specifically, exploration of newer architectures like Tensor Cores for factors like accuracy and computational complexity of channel estimation need to be addressed. 

\begin{enumerate}
    \item We implement a Correlation based Channel Estimation algorithm which uses PN sequences as Pilot signals. We use the auto-correlation properties of the PN sequences to design multiplexed Pilot transmission which can reduce the Pilot overhead while maintaining low error for the Channel Estimation.
    \item We then implement the Channel Estimation algorithm for Massive MIMO systems in a GPU and show the performance of Channel Estimation while varying factors such as PN sequence length, number of multiplexed pilots, Channel Impulse Response length, and Massive MIMO scale.
\end{enumerate}
\section{System Model}\label{sec:sys_model}

Consider an $N_t \times N_r$ MIMO system where $N_t$ and $N_r$ are the transmitting and receiving antennas respectively. Each transmitting antenna $t$ sends a Pilot sequence of length $P$ which is represented by vector $p_{t}$. We consider an i.i.d. Gaussian channel with maximum $L$ length CIR. Within the $L$ length channel there are $L_{nz} \leq L$ number of taps spread randomly with uniform distribution, and each tap is normalized such that maximum power of each tap is $\frac{1}{N_t\sqrt{L_{nz}}}$. Firstly, we consider the sequential transmission of pilots. To deal with the high pilot overhead for sequential transmissions, we then consider spatially multiplexed transmission case, where multiple transmitters send Pilot sequences simultaneously.

\subsection{Sequential Pilot transmission}\label{subsec:concur_tx}

\begin{figure}
    \centering
    \includegraphics[width=0.6\columnwidth]{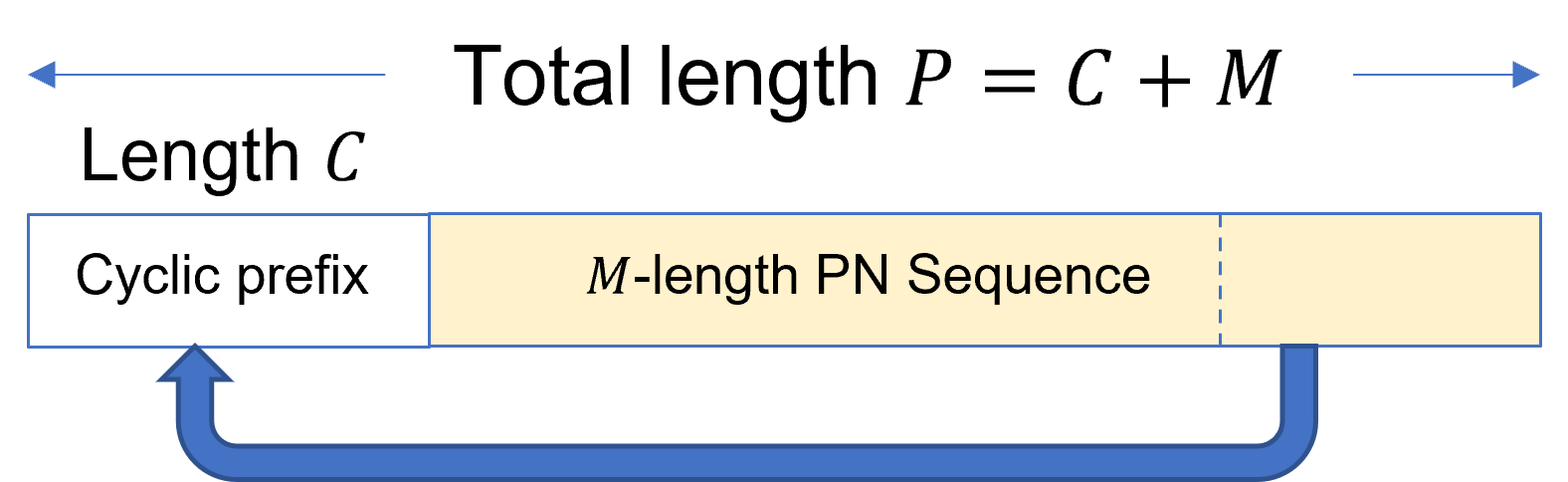}
    \caption{Pilot Sequence after adding Cyclic Prefix to the PN sequence}
    \label{fig:pilot_seq}
\end{figure}

\begin{figure}
    \centering
    \includegraphics[width=0.7\columnwidth]{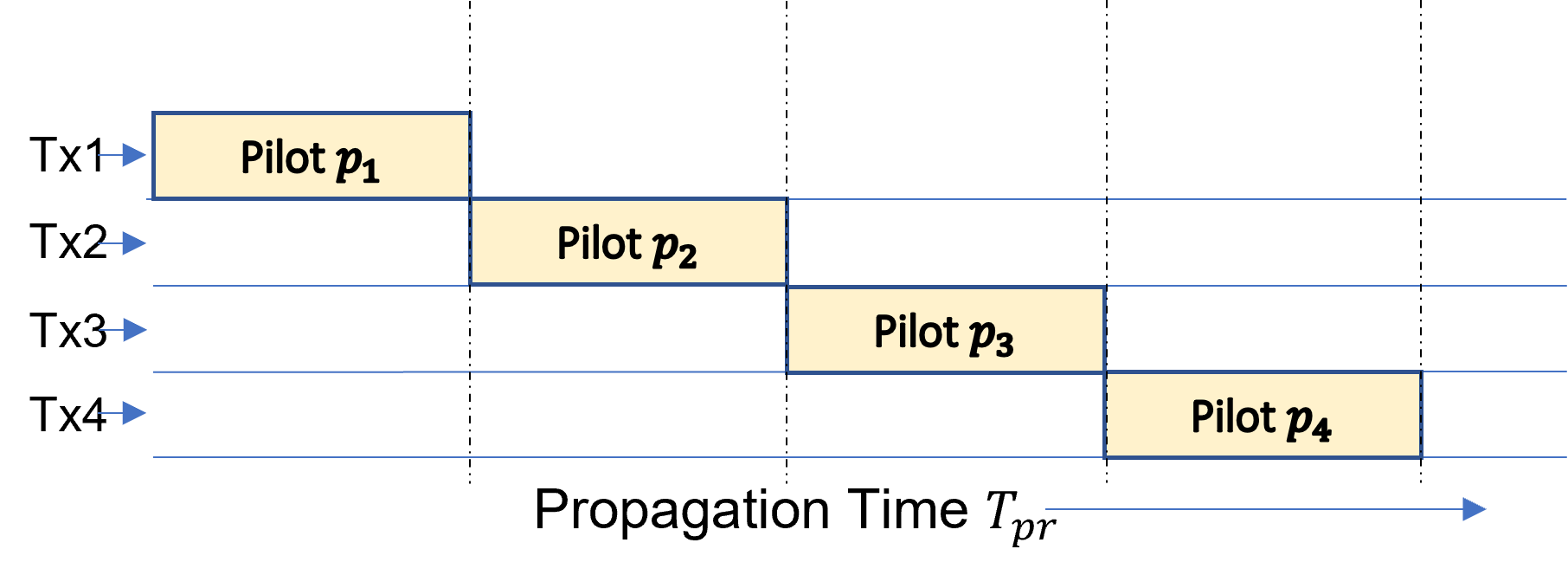}
    \caption{Sequential Pilot transmission}
    \label{fig:concur_pilot}
\end{figure}

We consider the use of an $M$-length PN sequence as Pilot signals. The circular auto-correlation properties of $M$-length PN sequence are well known and can be given as,
\begin{equation}\label{eq:pn_corr}
R[n] = \frac{1}{M}\sum_{m = 0}^{M-1} x_{t}[m] x_{t}^{*}[m + n] =
\begin{cases}
    1, & \text{if } n = 0 \\
    -\frac{1}{M}, & \text{if } 1 \leq n \leq M
\end{cases}
\end{equation}
Where $R[n]$ is the circular auto-correlation, and $*$ denotes the complex conjugate. The correlation in Eq. \ref{eq:pn_corr} approximates to the Delta function for long PN sequences. So, if we transmit PN sequences as pilots, we can then perform correlation of the received PN sequence with the transmitted PN sequence to get an estimated CIR.

To exploit this property, we create a Pilot sequence consisting of a $M$-length PN Sequence denoted by $s = [s_0, s_1,...,s_{M-1}]$, and a Cyclic Prefix (CP) of length $C$. This CP consists of the last $C$ samples of the PN sequence, and $L \leq C \leq M$ since we need at least $L$ samples to absorb the ISI (Inter-Symbol Interference) effect of the multipath channel. More importantly, the CP gives a periodic nature to the Pilot sequence. So, when transmitted through the channel, circular correlation can be performed on the received PN sequence to get the CIR. An illustration of the Pilot sequence of length $P = C + M$ is shown in Fig.\ref{fig:pilot_seq}. 

The Pilot sequences are passed through the $L$ length channel. An illustration of the sequential transmission case is shown in Fig. \ref{fig:concur_pilot}. After sequential pilot transmissions by all transmitting antennas, the received PN sequence can be shown in the form of matrix multiplication. The received signal at antenna $r$ for transmitting antenna $t$ will be,
\begin{equation}\label{eq:sys_model_matrix}
    Y_r = \mathbf{S_t}h_{r,t} + W
\end{equation}
where $Y_r$ is the $M \times 1$ vector at receiver $r$, $x_t$ is the Pilot sequence for transmitter $t$, $h_{r,t} = [\mathrm{h_0},\mathrm{h_0},...,\mathrm{h_{L-1}},0,..,0]^T$ is the $M$ length CIR. Here, within the $L$ indices, only $L_{nz}$ number of channel taps exist and rest of the values are zero. $\mathbf{S_t}$ is the matrix representation of the PN sequence in $M \times M$ circulant form i.e. first row of matrix is $[s_0, s_{M-1}, s_{M-2}, ..., s_{1}]$, second row is $[s_1, s_0, s_{M-1},...x_{2}]$ and so on, 
and $W$ is the Additive White Gaussian Noise (AWGN). Here, CP is not considered in Eq. \ref{eq:sys_model_matrix} as it represents the effect of CIR on the PN sequence.

We assume a real-valued PN sequence is transmitted. At the receiver, the CP is removed and only the PN sequence part of the received signal is taken for correlation. Due to the circular correlation property, the correlation based estimation is implemented as follows,
\begin{equation}\label{eq:correlator}
    \hat{H} = \frac{1}{M}S_t^TY_r = \frac{1}{M}S_t^T\mathbf{S_t}h_{r,t} + \frac{1}{M}S_t^TW
\end{equation}
For long PN sequences, due to noise mitigation and Eq. \ref{eq:pn_corr}, the channel estimate can then be given as,
\begin{equation}\label{eq:approx_chanest}
    \hat{H} \approx h_{r,t}
\end{equation}
Due to sequential transmissions, there will be no interference between transmitters. The total transmitted samples at each receiver are $P \times N_t$. If we denote the sampling rate by $F_s$, then the propagation time can be denoted as,
\begin{equation}\label{eq:prop_time}
    T_{pr} = \frac{P \times N_t}{F_s}
\end{equation}
Intuitively, for the sequential transmission case, as $N_t$ increases, due to the increase in Propagation time, the channel estimation overhead increases. Transmission time can be decreased by reducing PN sequence length. But this results in increase in the estimation error.

\subsection{Multiplexed Pilot transmission}\label{subsec:multi_tx}

\begin{figure}
    \centering
    \includegraphics[width=0.7\columnwidth]{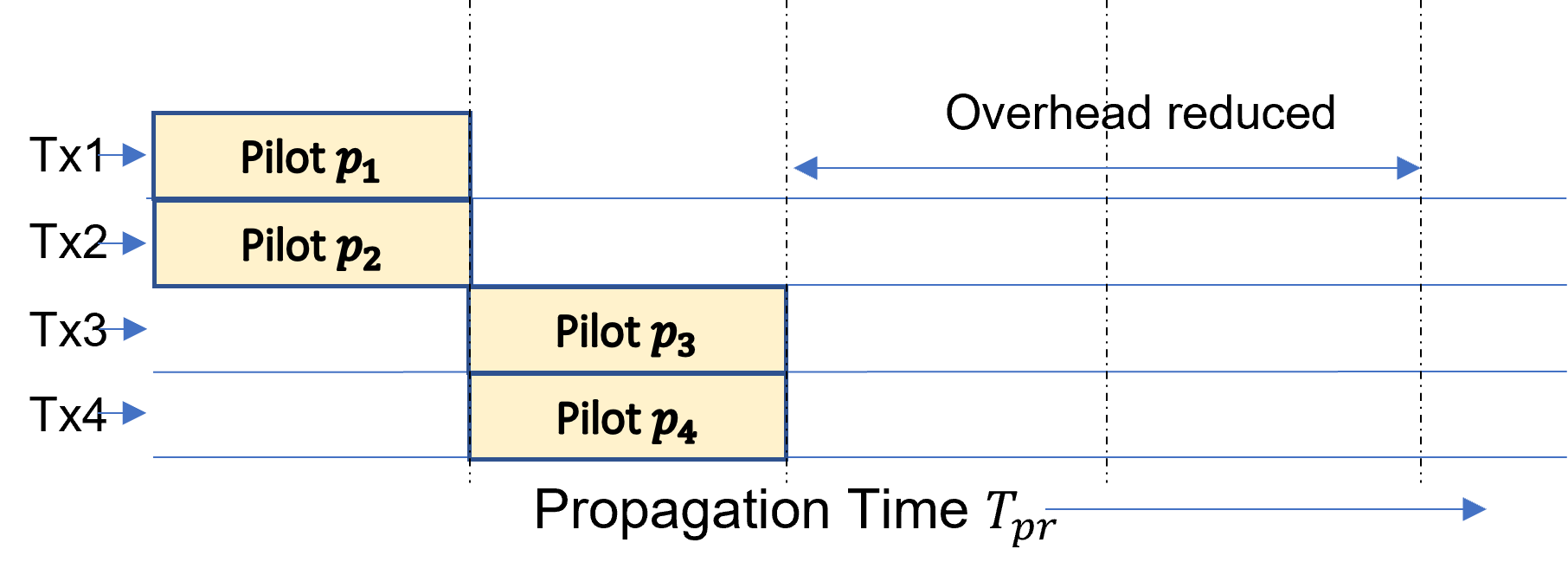}
    \caption{Multiplexed pilot transmission with $N_{batch} = 2$.}
    \label{fig:batch_tx}
\end{figure}

To reduce the channel estimation overhead, simultaneous transmissions can be performed by multiple transmitters. We assume $N_{batch}$ transmitters will be sending Pilot sequences simultaneously, where $1 \leq N_{batch} \leq N_t$. The received signal, without the CP, at each antenna $r$ can be given as,
\begin{equation}\label{eq:sys_model_with_int}
    Y_r = \mathbf{S_t}h_{r,t} + \sum_{n_i = 1}^{N_{batch}}\mathbf{S_{n_i}}h_{r,n_i} + W
\end{equation}
Where $\mathbf{S_{n_i}}$ is the circulant matrix representation of the PN sequence, and $h_{r,n_i}$ is the CIR for transmitting antenna $n_i$.

Since multiple transmitters send Pilot sequences simultaneously, for a transmitter $t$, all other simultaneously transmitter sequences will be interference. To mitigate this, an option is to use multiple PN sequences. But using multiple PN sequences can become an issue since the number of PN sequences with minimal cross-correlation are very limited. Also, the cross-correlation between PN sequences will not approximate to Eq. \ref{eq:pn_corr}, which means interference will not be mitigated completely. Instead, we use a single PN sequence with circular shifts for multiple antennas which are transmitting simultaneously.

The assumption made here is that CIR length $L << M$, which means correlating with only $L$ circular shifts is required to get the CIR. Circular shifted row multiplications in Eq. \ref{eq:correlator} beyond the CIR length between matrix $S_t$ and $Y_r$ become redundant as no new information is obtained. This assumption also holds true for the sequential transmission case since CIR estimate with low error can only be obtained with longer PN sequences. So, if we circularly shift the PN sequences beyond the CIR length, they can be simultaneously transmitted without interference between the sequences.

To design such Pilot sequences, we consider the $P$ length Pilots used in Sec. \ref{subsec:concur_tx}. We then batch transmitters depending on how many circularly shifted PN sequences can be obtained. This is determined by the $C$ length CP. We assume to have information about the maximum CIR length $L$ and consider $C = L$. So, the number of transmitters that can be batched are,
\begin{equation}\label{eq:num_batch}
    N_{batch} \leq floor \left(\frac{M}{C} \right)
\end{equation}
The $N_{batch}$ transmitters are then given a circularly shifted version of the same PN sequence where the circular shift for transmitter $t$ is given by,
\begin{equation}\label{eq:circ_shift}
    C_{shift} = \frac{M}{N_{batch}} \times (t \text{ mod } N_{batch})
\end{equation}
After the Pilot sequences are circularly shifted, the CP is added by taking the last $C$ samples of the circularly shifted PN sequence after which multiplexed transmissions are performed. An illustration of multiplexed transmissions is shown in Fig. \ref{fig:batch_tx}. The Pilot sequences received can be given as,
\begin{multline}
    Y_r^b = [(S_{1}^{p})^T:(S_{2}^{p})^T: ...:(S_{N_{batch}-1}^{p})^T]^T \\ [h_{r,1}^T:h_{r,2}^T:...:h_{r,N_{batch}-1}^T]^T + W
\end{multline}\label{eq:sys_model_batched}
where $S_{i}^{p}$ is the $\frac{M}{N_{batch}} \times P$ partial circulant matrix representation of the circularly shifted Pilot sequence. Removing the CP and performing circular correlation, we get,
\begin{equation}\label{eq:batched_cir}
    \hat{H}_{batch} = \frac{1}{N}S_tY_r^b \\ \approx [h_{r,0}^T:h_{r,1}^T:...:h_{r,N_{batch} - 1}^T] 
\end{equation}
for PN sequence length $>> L$. Here $\hat{H}_{batch}$ is the CIR estimate consisting of CIRs of all transmitters whose Pilot sequences were sent in the same batch, So, the multiplexed transmission case has a propagation time of,
\begin{equation}\label{eq:batch_prop_time}
    T_{pr}^b = \frac{P}{F_s} \times \frac{N_t}{N_{batch}}
\end{equation}
which shows a reduction of the Pilot overhead by $N_{batch}$ when compared with Eq. \ref{eq:prop_time}. While the multiplexed transmission case gives overhead reduction, an interesting thing to note is the potential to help mitigate the Pilot Contamination issue that occurs in Multi-User multi-cell Massive MIMO systems. Since only a single PN sequence is being used for all transmitters, other $M$-length PN sequences can be used by neighboring cells for channel estimation with mitigated interference.
\section{Implementation}\label{sec:impl}

\subsection{Setup}\label{subsec:setup}

Consider the $N_t \times N_r$ MIMO system from Sec. \ref{sec:sys_model}. For our implementation scenario we assume both $N_t$ and $N_r$ are $\geq 16$ and an $L$ length channel with $L_{nz}$ taps. We simulate the Pilot sequence transmission through such the channel and then reception at the $N_r$ receiving antennas. We consider the receiver side array connected to a server consisting of CPU and GPU via a switch. Here the CPU is used for control and data access whereas the GPU is used for the bulk of data processing. This type of setup has been used in various testbed deployments such as ORBIT \cite{raychaudhuri2005overview} and COSMOS \cite{raychaudhuri2020challenge} due to the modular structure and flexibility that it provides for implementing various scenarios and testing multiple algorithms. We use this setup present in a Sandbox in the ORBIT Testbed to simulate and implement the channel estimation scenarios. 

\subsection{Implementation in Tensor Cores}\label{subsec:tensor_core_impl}

Once the received sequences are sent to the GPUs for processing, we perform the correlation by using the Tensor Cores present in each GPU. For this implementation, we use the Nvidia V100 GPUs which consists of $5120$ CUDA Cores and $640$ Tensor Cores, $1530$ MHz clock and $16$ Gb of DRAM \cite{nvidiavoltawhitepaper}. CUDA architecture abstraction in combination with the Tensor Cores is used for programming the correlation in the GPU \cite{markidis2018nvidia}.

Tensor Cores are being included in modern Nvidia (R) GPUs as an alternative micro-architecture to the CUDA cores. Tensor cores are specialized micro-architectures which will perform parallel $4 \times 4$ matrix-multiply and accumulate operation for half-precision (16-bit) floating point numbers \cite{markidis2018nvidia}. Since half-precision floats are used, the data which is usually in full-precision complex floats, is converted down to half-precision. This is done in combination with the CP removal before correlation is to be performed in the GPU.

The correlation operation, as shown in Eqs. \ref{eq:correlator} and \ref{eq:batched_cir}, can be considered as a matrix multiplication of the circulant matrix representation of the PN sequence and the received signal. So, we can utilize these Tensor Cores present in the GPUs to implement the correlation based channel estimation as a parallel matrix multiplication. Since multiple Tensor Cores exist in the GPUs, each core can take a smaller $4 \times 4$ tile of the matrix multiplication to be performed, and gives a $4 \times 4$ multiplication output which is stored in an output matrix.

While Tensor cores can provide fast processing, since they use half-precision arithmetic, the range of numeric representation can become very low. So, the final multiplication output can have increased error due to the rounding errors and early saturation of correlation output. 
We mitigate these errors occurring due to low precision computations by performing the normalizing operation i.e. $\left( \frac{1}{M} \right)$ in Eq. \ref{eq:batched_cir}, of the correlation after partial multiplication instead of normalizing after the complete matrix multiplication. This keeps the multiplication output below the saturation point for half-precision computation. 
\section{Experiments and Results}\label{sec:exp_and_res}

\begin{table}[]
    \centering
    \begin{tabular}{|c|c|}
        \hline
        Parameter & Value \\
        \hline
        \hline
        MIMO scale & $16 \times 16$,$32 \times 32$,$64 \times 64$ \\
        \hline
        PN Sequence Length & 511, 1023, 2047 \\
        \hline
        CP length & 64 - 128 \\
        \hline
        No. of Channel Taps & 1 - 128 \\
        \hline
        $N_{batch}$ & 1,2,4,8 \\
        \hline
        Iterations per experiment & 50 \\
        \hline
    \end{tabular}
    \caption{Parameter values used for various experiments performed}
    \label{tab:exp_params}
\end{table}

Table \ref{tab:exp_params} shows parameter values that will be used when performing experiments. The crucial goal is to evaluate the performance of multiplexed PN sequence transmissions, and then investigate the processing latency and error performance when using Tensor Cores to perform channel estimation. We firstly look into the Mean Absolute Error (MAE) performance of the Channel Estimation. We define MAE as follows,
\begin{equation}\label{eq:mse}
    E = \frac{1}{LN_tN_r}\sum_{t = 1}^{N_t}\sum_{r = 1}^{N_r}\sum_{l = 1}^{L}|\hat{h}_{r,t}[l] - h_{r,t}[l]|
\end{equation}
where $|.|$ denotes the absolute value.

\begin{figure}[t]
    \centering
    \includegraphics[width=0.8\columnwidth]{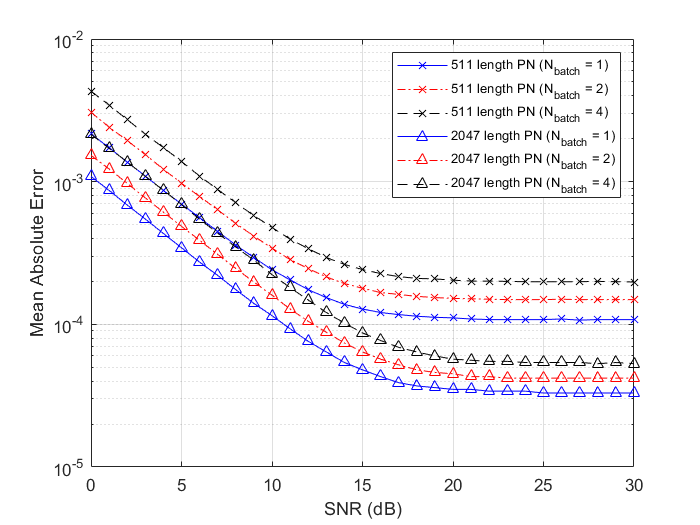}
    \caption{MAE vs SNR for PN length $M = 511,2047$ and $N_{batch} = 1,2,4$. $L = 64, L_{nz} = 64, N_r = N_t = 16$}
    \label{fig:mse_for_pn_seq_var}
\end{figure}

\subsection{Mean Absolute Error vs SNR for different PN sequence lengths}

We keep the channel length $L = 64$ and channel taps $L_{nz}=64$. Fig. \ref{fig:mse_for_pn_seq_var} shows the MAE vs SNR graph for various PN sequence lengths and $N_{batch}$ values. The PN sequence length affects the MAE performance significantly due to the correlation of longer PN sequences being close to the Delta function. While lower SNR regions affect the MAE, an error floor is seen at higher SNRs which reduces with increase in PN sequence length. The error floor mainly exists due to the $-\frac{1}{M}$ in Eq. \ref{eq:pn_corr}. But the Tensor Core implementation also introduces errors due to low precision. Also, variation in the pilot multiplexing can vary the MAE, but the MAE difference remains consistent throughout the SNR range. 

\subsection{Mean Absolute Error vs SNR for different number of channel taps}

\begin{figure}[t]
    \centering
    \includegraphics[width=0.8\columnwidth]{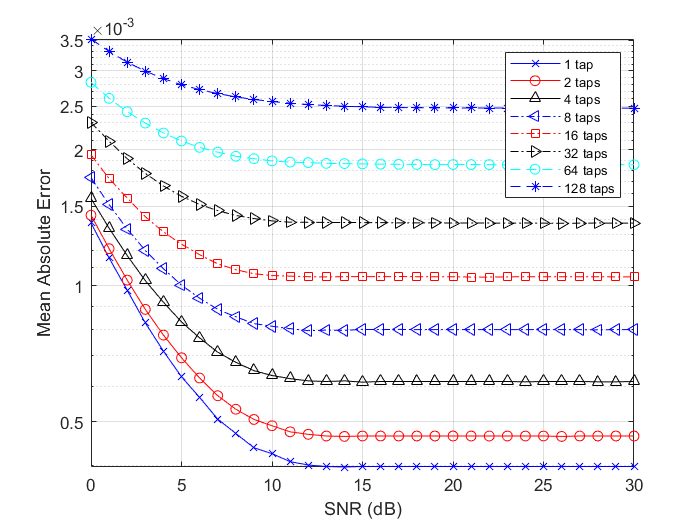}
    \caption{MAE vs SNR for channel estimation while varying number of channel taps. PN length $M = 2047,N_{batch} = 4,L = 128,N_t = N_r = 16$}
    \label{fig:mse_for_chan_taps}
\end{figure}

To look closely at the error performance due to use of Tensor Cores, we perform an experiment to measure the MAE while varying the SNR and the channel taps for a $2047$-length PN sequence. The goal for this experiment is to evaluate the Tensor Core implementation accuracy when performing correlation for varying channel taps. $4$ transmitter are multiplexed, the channel length $L = 128$, but the number of channel taps within channel length vary from $1$ to $128$. Fig. \ref{fig:mse_for_chan_taps} shows the effect of varying the channel taps. $L$ length correlation is still performed by the Tensor Cores, as the number of channel taps change within the channel length, number of near non-zero correlated values will change. This affects the Tensor Core performance since the low precision computation could give increased errors for high number of channel taps. This is seen in Fig. \ref{fig:mse_for_chan_taps}, where PN sequence length is unchanged but the MAE is high for the $L_{nz} = 128$, and low for small number of $L_{nz}$. Here, since the channel length does not change which means correlation length is unchanged. So the processing latency will not be affected.

\begin{figure}[t]
    \centering
    \includegraphics[width=0.8\columnwidth]{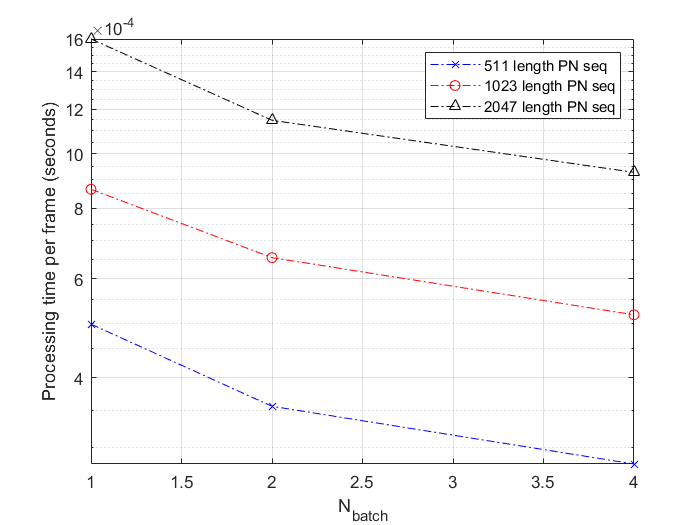}
    \caption{Processing latency vs $N_{batch}$ value for various PN sequence lengths. $L=64,L_{nz}=64,N_t = N_r = 16$}
    \label{fig:proc_lat_vs_batch_pn}
\end{figure}

\subsection{Frame Processing Latency vs number of multiplexed antennas for different PN sequence lengths}

We then investigate the processing latency for the Tensor core implementation for a one frame. Here, a frame consists of Pilot transmissions for all transmitters and its length is $\frac{P*N_{t}}{N_{batch}}$. Correlation is performed for various PN sequence lengths and various $N_{batch}$ values. For calculation of the processing time, the correlation time as well as the memory transfer time between CPU and GPU is considered. From Fig. \ref{fig:proc_lat_vs_batch_pn}, it can be seen that the processing latency is heavily affected by the multiplexed pilot transmissions. This is mainly due to decrease in frame size when pilots are multiplexed, leading to decrease in memory transfer time between the CPU and GPU. Also, when we compare the processing latency with the frame propagation time from Eq. \ref{eq:batch_prop_time}, the Tensor Core processing time can be lower than or equivalent to the propagation time for sampling rates up to $10$ MHz. This makes it possible to use the implementation in real-time Massive MIMO deployments. 

\begin{figure}[t]
    \centering
    \includegraphics[width=0.8\columnwidth]{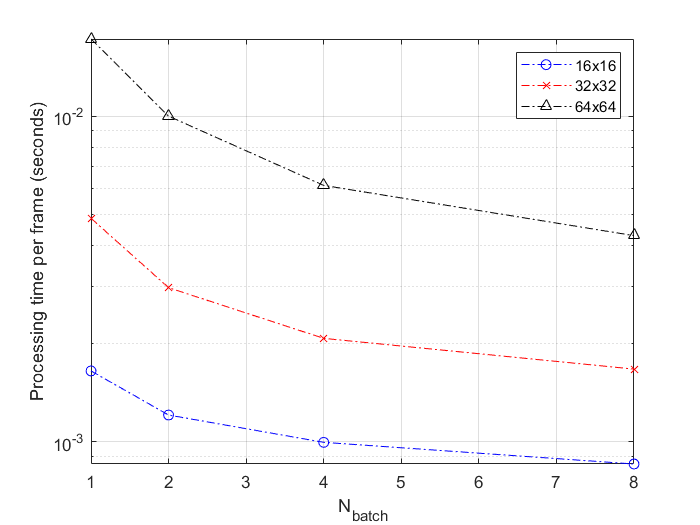}
    \caption{Processing latency vs $N_{batch}$ value for variation in Massive MIMO array size. $M = 2047, L = 128, L_{nz} = 20$}
    \label{fig:proc_lat_vs_batch_array_size}
\end{figure}

\subsection{Frame Processing Latency vs number of multiplexed antennas for different MIMO array sizes}

We increase the scale of antenna arrays and evaluate the processing latency of the correlation. Fig. \ref{fig:proc_lat_vs_batch_array_size} shows the variation in Processing time per frame vs the $N_{batch}$ value for various Massive MIMO scales. We consider the $32\times32$ and $64\times 64$ MIMO systems and compare the Frame processing latency. With no multiplexing, the processing latency varies highly with the antenna array size. But when high multiplexing is performed, the effect reduces. This effect is similarly seen in the PN sequence variation case in Fig. \ref{fig:proc_lat_vs_batch_pn}, where the decrease in memory transfer time decreases the overall processing latency. This becomes more apparent for larger MIMO scale where the frame size also increases considerably. 

\section{Conclusion}\label{sec:conc}

We implemented a fast and low-overhead channel estimation scheme for Massive MIMO systems in a GPU by exploiting the PN sequence correlation properties and using multiplexed pilot transmissions. To deal with the high complexity of correlation based channel estimation, we implemented the channel estimation by using Tensor Cores in Nvidia GPUs. We then performed experiments for various scenarios to evaluate the error performance and estimation accuracy of the Tensor Core based implementation.

While Tensor Core implementation shows potential, using only single GPU for processing limits the scale at which Massive MIMO systems can be used. To deal with this issue, further investigation into distributed multi-GPU processing is required. Heterogeneous processing architectures which combine CPUs and GPUs for processing should also be investigated.

\bibliographystyle{ieeetr}
\bibliography{ref.bib}

\end{document}